\documentclass[aps,prd,reprint,superscriptaddress,nofootinbib,floatfix]{revtex4-2}
\usepackage[utf8]{inputenc}
\usepackage[T1]{fontenc} 

\usepackage{amsmath}
\usepackage{amssymb}
\allowdisplaybreaks

\usepackage{graphicx}
\usepackage{dcolumn}
\usepackage{xcolor}
\usepackage{float}
\usepackage{bm}
\usepackage{comment}
\usepackage{tabularx}
\usepackage{longtable}
\usepackage{booktabs}

\usepackage{microtype}
\usepackage[italicdiff]{physics}
\usepackage[colorlinks=True,allcolors=blue]{hyperref}

\newcommand\IITRxPH{Department of Physics,
Indian Institute of Technology Roorkee, 
Roorkee 247667, India} 

\newcommand\UGxICTQT{International Centre for Theory of Quantum Technologies,
University of Gda\'{n}sk,
80-308 Gda\'{n}sk, Poland}

\newcommand\XMUMxPH{School of Mathematics and Physics, 
Xiamen University Malaysia,
43900 Sepang, Malaysia}

\newcommand\UGxIFTiA{Institute of Theoretical Physics and Astrophysics,
Faculty of Mathematics, Physics and Informatics,
University of Gda\'{n}sk,
80-308 Gda\'{n}sk, Poland}

\newcommand\IITRxCPQCT{Centre for Photonics and Quantum Communication Technology, 
Indian Institute of Technology Roorkee, Roorkee 247667, 
India}

\begin{document}

\title{Probing Modified Gravity with Entanglement of Microspheres}

\author{Ankit Kumar}
\email[Ankit Kumar: ]{kumar.ankit.vyas@gmail.com}
\affiliation{\IITRxPH}
\affiliation{\UGxICTQT}

\author{Yen-Kheng Lim}
\affiliation{\XMUMxPH}

\author{P. Arumugam}
\affiliation{\IITRxPH}
\affiliation{\IITRxCPQCT}

\author{Tom Z{\l}o\'{s}nik}
\affiliation{\UGxIFTiA}

\author{Tomasz Paterek}
\affiliation{\XMUMxPH}
\affiliation{\UGxIFTiA}

\begin{abstract}
While a wide variety of astrophysical and cosmological phenomena suggest the presence of Dark Matter, all evidence remains via its gravitational effect on the known matter. As such, it is conceivable that this evidence could be explained by a modification to gravitation and/or concepts of inertia. Various formulations of modified gravity exist, each giving rise to several non-canonical outcomes. This motivates us to propose an experiment searching for departures from (quantum) Newtonian predictions in a bipartite setting with gravitational accelerations $\lesssim 10^{-10}$ m/s$^2$, i.e., where the effective force needs to be stronger than Newtonian to account for the Dark Matter effects. Since quantum particles naturally source weak gravitation, their non-relativistic dynamics offers opportunities to test this small acceleration regime. We show that two nearby mesoscopic quantum masses accumulate significantly larger entanglement in modified gravity models, such as the Modified Newtonian Dynamics. 
Our calculations include Casimir-Polder forces as well as tidal effects next to the surface of the earth, and confirm that entanglement is observable within the limits imposed by environmental decoherence.
We demonstrate how the temperature can be fine-tuned such that modified gravity is certified simply by witnessing the entanglement generated from uncorrelated thermal states, eliminating the need for precise noise characterization.
Overall, the required parameters could be realized in a tabletop experiment.
\end{abstract}

\maketitle

\emph{\textbf{Introduction.}}
The Newtonian limit of General Relativity is very successful on the scale of the solar system.
For example, balancing the centrifugal and the gravitational forces of objects in approximately circular orbits around the Sun implies that orbital velocity falls as the square root of the distance. 
This is famously known as Keplerian decline and has been observed to hold for all planets~\cite{NASA-PlanetFactSheet-rel2Earth}.
Spiral galaxies have a lot in common with the solar system. 
Most of their mass is also concentrated towards the center, but the stars do not show any asymptotic Keplerian decline ~\cite{book-DM-Sanders2010}. 
Their orbital speeds generally do not fall, and the rotation curves saturate~\cite{Andromeda-Babock1939}.
Consequently, the stars in the outer regions appear to be orbiting so fast that they should not be gravitationally bound.
This is not happening, and hence there seems to be more gravity than expected based on the known visible mass at the center of spiral galaxies. 
This is a prime example of the Dark Matter (DM) effect. The name originates in the proposal of the existence of an invisible matter distributed throughout the galaxies~\cite{ComaCluster-Zwicky1937}, generating an extra gravitational pull that balances the centrifugal force.
Despite being the most widely accepted explanation and with evidence appearing even on the largest cosmological scales, existence of DM has not been directly detected or confirmed by any experiment so far~\cite{DM-Review-Bertone2018}, and hence the continued interest in alternative solutions.
A plausible route involves modifications to our present understanding of gravity.
Modified Newtonian Dynamics (MOND) is one such proposal where, without invoking DM, Newton's second law and/or the law of universal gravitation is modified to account for DM effects in galaxies~\cite{MOND-Milogram1983}.
While the experiment we propose is independent of any concrete formulation of alternative gravity models, for quantitative statements we follow the parameters present in the MOND theory. We therefore begin with more details about it.

\emph{\textbf{Modified Newtonian Dynamics.}}
A general form of MOND, which encompasses a wide variety of proposed variants of the model~\cite{MOND-Milgrom1992,Lag4MOND-Bekenstein1984,TeVeS-Bekenstein2004,MOND-QLinear-Milgrom2009,Milgrom:2023idw,MOND-General-Milgrom2023}
, generalizes Newton's second law to 
\begin{equation}
\nu\bigg(\frac{|\vec{a}|}{a_{0}}\bigg)\vec{a} = -\nabla\Phi + \vec{F}/m,
\label{eq:DiffMOND-1}
\end{equation}
where $\vec{F}$ represents the sum of all non-gravitational forces on an object of mass $m$, $\vec a$ is its acceleration, and $a_0$ is the acceleration scale where the generalization is to take place. The potential $\Phi$ belongs to the set of potentials $\{ \phi_{i} \}$, or it could be their linear combination.
Collectively, the potentials obey a system of potentially nonlinear and coupled Poisson equations:
\begin{equation}
\sum_{j}\nabla \cdot\bigg[\mu_{ij}\bigg(\frac{{\nabla}\phi_{1}}{a_{0}}, \frac{{\nabla}\phi_{2}}{a_{0}}, \dots \bigg) \nabla \phi_{j}\bigg] = 4\pi G_{i} \rho,
\label{eq:DiffMOND-2}
\end{equation}
where $\rho$ is the density of matter, $G_{i}$ is the generalized Newton's constant quantifying the coupling of matter with the potential $\phi_{i}$, and 
$(\nu,\mu_{ij})$ are the interpolating functions that govern the transition from Newtonian dynamics to the modified regime. Many existing models are special cases of the above expressions (see Supplemental Material for various existing formulations). 
Proposals where $d\nu(x)/dx \neq 0$ are referred to as models of \emph{modified inertia}.
It is not known whether a viable action principle for these models exists, and so it is not clear whether the degrees of freedom describing an object can be canonically quantized.
Since we explore the possibility of entanglement accumulation as the signal revealing non-canonical gravity, we effectively put to test only the \emph{modified gravity} formulations of MOND ($\nu(x)=1$), as the existence of a Hamiltonian allows us to employ standard quantum mechanical methods. The exact functional forms for $\mu_{ij}$ are still debated, but consistency with astronomical observations fixes their asymptotics.
For example, in the Lagrangian-based AQUAL model $\nu=1$ and there exists a single potential $\Phi$ with interpolating function $\tilde{\mu}$~\cite{Lag4MOND-Bekenstein1984}. 
In the limiting cases $\tilde \mu(x\ll 1) \to x$ to explain galactic rotation curves, and $ \tilde \mu(x\gg 1) \to 1$ to recover Newtonian dynamics in regimes of stronger gravity.
Note that even a single massive object may source multiple potentials $\phi_{i}$.
Notably, although each model is non-relativistic, some exist as limits of relativistic completions. 
A recent example of such a completion, building on antecedent work~\cite{TeVeS-Bekenstein2004}, was shown to be consistent with data from the large-scale cosmic structure and inhomogeneities of the cosmic microwave background~\cite{RMOND-Zlosnik2021}.
Although there does not exist a unique model for the MOND paradigm, we will see that many of them and potentially other modified gravity theories predict deviations from Newtonian gravity that may be observable in bipartite quantum mechanical experiments.

A wide array of DM phenomenology in galaxies is associated with acceleration due to gravity below $a_0 \approx 1.2 \times 10^{-10}$ m/s$^2$~\cite{MOND-Milogram1983,MONDinMilkyWay-Famaey2005,MOND-Gentile2011}. 
Such small accelerations are readily accessible between quantum masses.
We show here how the entanglement dynamics between two microspheres senses the force gradient between them, and design tests where a simple act of entanglement witnessing reveals non-canonical interaction. 
The calculations include Casimir-Polder forces, tidal effects next to the surface of the earth, and estimates for the effects of environmental decoherence.
The modified gravity models have a stronger force gradient than Newtonian in the discussed limit, thereby leading to a stronger entanglement that is robust against noisy measurements.
Such an experiment could be done using the masses recently cooled in Vienna~\cite{Sciene-Aspelmeyer2020} which, when separated by a distance of a few times their radius, generate an acceleration deep into the discussed relevant regime.
Our calculations at first assume the particles are in free space devoid of all external gravitational fields.
Although this might seem unrealistic, there may be places in the solar system where the net gravitational acceleration is in the MOND regime~\cite{MOND-Bekenstein2006,MOND-LISA-Magueijo2012}.
Furthermore, it is also instructive to perform this experiment either with freely falling microspheres in a large orbit around the earth, where pressure falls to $\sim 10^{-15}$ Pa and the decoherence effects are negligible, or in a laboratory on earth. In the latter case, the acceleration along the line between the masses, sourced by the tidal force, is two orders of magnitude larger than the acceleration due to mutual gravity. Yet, we show that it has a negligible effect on generated entanglement and hence opens the way towards a terrestrial experiment.
An observation of non-canonical effects in agreement with our calculations would indicate that the modified gravity adheres to the Strong Equivalence Principle (SEP), i.e., the internal dynamics stays the same in all uniformly accelerated frames.
See also Ref.~\cite{MOND-SEPwSTEP-Pereira2016}, which discusses probing MOND with the Satellite Test of the Equivalence Principle. Note that SEP is certainly violated in the MOND model.
Many other proposals have been put forward to test the predictions of MOND. Torsion pendula experiments show an agreement with Newton's second law down to accelerations $\approx a_0/2400$~\cite{NewtonLaw-TorsionPend-Gundlach2007}, and with Newton's law of universal gravity down to $\approx a_0/60$~\cite{TestNewton-Little2014}.
The accelerations mentioned in these experiments are with respect to the local laboratory, and the net acceleration due to heavenly sources is well above $a_0$. Therefore, it remains unclear to what extent departures from Newtonian dynamics are to be expected.

The driving idea behind modified gravity models is the following: just like Newtonian gravity is an approximation of General Relativity when the gravitational field is not too strong, it might also be an approximation of an underlying theory when the Newtonian accelerations are not too small.
In order to match the stellar rotation curves, the modification has to involve accelerations smaller than $a_0 \approx 1.2 \times 10^{-10}$ m/s$^2$, and the simplest solution is to demand that gravitational force in this regime scales inversely to the distance.
Accordingly, in the deep MOND regime the modified gravitational potential at a position $\vec r_2$ due to a mass $m$ located at position $\vec r_1$ is given by $\Phi = \sqrt{Gma_0} \ \ln( \abs{\vec r_2 - \vec r_1})$. 
Note that this violates the law of equal and opposite action and reaction: the force on a mass $m_2$ due to $m_1$ is $\sim m_2\sqrt{m_1}$, whereas the force on mass $m_1$ due to $m_2$ is $\sim m_1\sqrt{m_2}$.
This issue is rectified if one solves the full nonlinear Poisson-like equation governing the bipartite dynamics in MONDian gravity.
The resultant gravitational potential energy of two identical particles of mass $m$ is given by~\cite{MOND2Body-Milgrom2014,MOND2Body-Zhao2010,MOND2Body-Milgrom1986}:
\begin{equation}
V_M(\vec r_1, \vec r_2) =  \frac{4}{3} \qty( \sqrt{2} - 1 ) \ m \sqrt{G m a_0} \ \ln(L+r) ,
\label{eq:ForceIdenticalMasses}    
\end{equation}
where $r$ 
is the \emph{relative displacement} of the two masses from their initial separation $L$.
This is very different from the usual Newtonian potential and underlies the differences in observable quantities.
In particular, we focus on quantum entanglement between small objects.

\emph{\textbf{Entanglement Dynamics.}}
\begin{figure*}
\centering
\includegraphics[width=\linewidth]{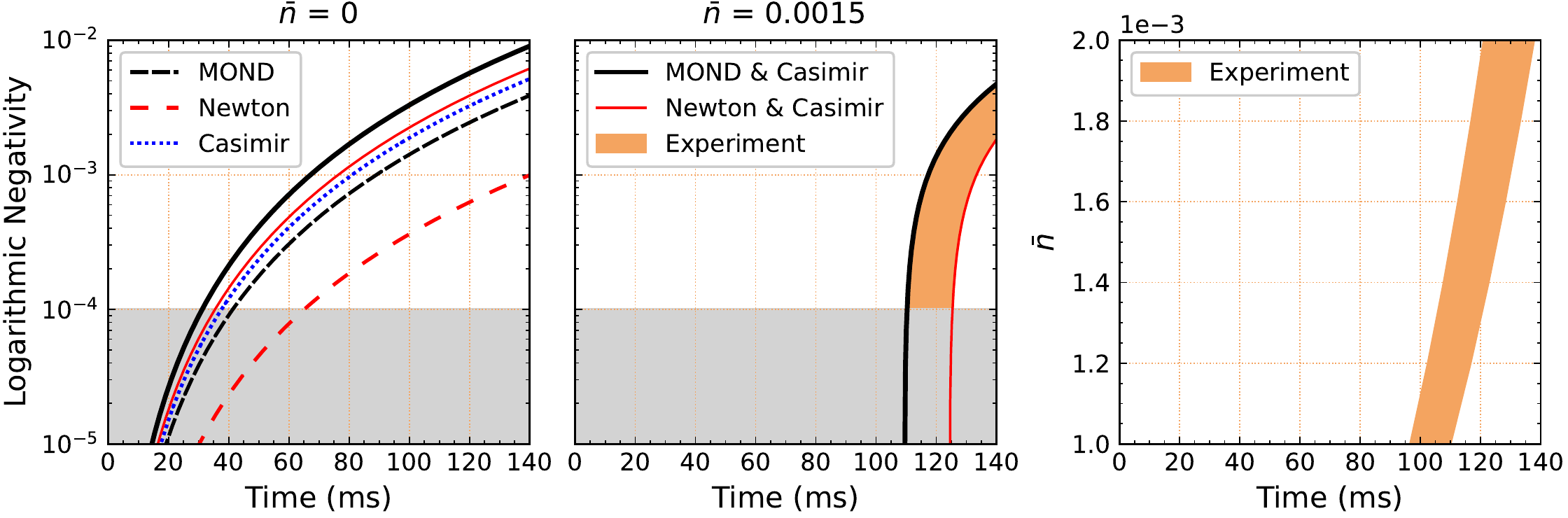}
\caption{Gravitational and Casimir entanglement between two identical Platinum spheres of radius 10 $\mu$m. The two masses are initially separated by a distance of 100 $\mu$m, and their initial quantum states are prepared by cooling in harmonic traps of frequency $10$ MHz. Entanglement is measured by logarithmic negativity (note also logarithmic vertical scale), $\bar{n}$ is the average phonon number (temperature),
and the grey shaded region (logarithmic negativity $< 10^{-4}$) signifies entanglement sensitivity beyond the limits achievable with near-future technology~\cite{Science-Palomaki2013,NeuroComp-Tanjung2023}. The dashed lines present calculations with only Newtonian potential, only gravity due to Modified Newtonian Dynamics (MOND) and taking into account only Casimir attraction. The solid lines describe the simultaneous presence of (modified) gravity and Casimir. The regions marked as ``Experiment'' correspond to entanglement values indicating modified gravity. Detecting any entanglement before the time when the red curve comes out of the grey region provides the simplest check of non-canonical interaction.}
\label{fig:Entanglement}
\end{figure*}
Entanglement dynamics of two nearby quantum masses in empty space, starting with uncorrelated thermal states, has been studied in detail in  Refs.~\cite{QuantGrav-Tanjung2020,QuantGrav-MesEnt-Qvarfort2020,QuantGrav-Ankit2022}. The system gets entangled because of the position dependence of the gravitational force: the parts of the wave packet closer to each other are attracted more than the parts further away. Accordingly, different momenta are generated at different positions, producing entanglement as time passes.
Notably, the methods of Ref.~\cite{QuantGrav-Ankit2022} apply to any central interactions without the need for an explicit form of the Hamiltonian.
The displacement-to-separation ratio $r/L$ is always small in gravitational experiments, and hence we expand Eq.~\eqref{eq:ForceIdenticalMasses} in a
Maclaurin series. Note that the cubic term is relevant in entanglement dynamics only when there is a significant relative motion between the two particles~\cite{QuantGrav-Ankit2022}, and hence we truncate at the quadratic term:
\begin{equation}
V_M(r)  \approx  \frac{4}{3} \qty( \sqrt{2} - 1 ) \ m \sqrt{Gma_0} \qty( \ln(L) +  \frac{r}{L} - \frac{r^2}{2L^2} ).
\end{equation}
For quantum particles this interaction is quantized by replacing variable $r$ with operator $\hat r$.
The system admits Gaussianity at all times, and we quantify entanglement through the logarithmic negativity of the bipartite covariance matrix~\cite{PPTCrit-CV-Simon2000,Negativity-Gaussian-Adesso2005,Negativity-Vidal2002,Negativity-CV-Serafini2004}.
The covariance matrix starting from the uncorrelated ground states of individual harmonic traps, each with frequency $\omega_0$, is derived in exact closed form for any quadratic and central potential in Ref.~\cite{QuantGrav-Ankit2022}, and it reads:
\begin{eqnarray}
\bm{\sigma}_{00}  &=& \frac{\hbar}{4m\omega_0} \qty[ 2+\omega_0^2t^2+\qty( 1+\frac{\omega_0^2}{\omega^2} )\sinh^2(\omega t) ],   
\nonumber \\
\bm{\sigma}_{02}  &=& \frac{\hbar}{4m\omega_0} \qty[ \omega_0^2t^2-\qty(1+\frac{\omega_0^2}{\omega^2})\sinh^2(\omega t) ],
\nonumber \\
\bm{\sigma}_{11}   &=&  \frac{m\hbar\omega_0}{4} \qty[ 2+\qty( 1+\frac{\omega^2}{\omega_0^2} )\sinh^2(\omega t) ], 
\nonumber \\
\bm{\sigma}_{13} &=& - \frac{m\hbar\omega_0}{4} \qty( 1+\frac{\omega^2}{\omega_0^2} ) \sinh^2(\omega t), 
\nonumber \\
\bm{\sigma}_{01} &=& \frac{\hbar}{8} \qty[  2\omega_0 t + \qty( \frac{\omega_0}{\omega}+\frac{\omega}{\omega_0} )\sinh(2\omega t) ],   
\nonumber \\
\bm{\sigma}_{03} &=& \frac{\hbar}{8} \qty[  2\omega_0 t - \qty( \frac{\omega_0}{\omega}+\frac{\omega}{\omega_0} )\sinh(2\omega t) ],
\label{eq:CovMat_FreeFall}
\end{eqnarray}
where $\omega$ encodes the mutual interaction between the two particles as follows.
For an arbitrary central potential expanded in a series in terms of the displacement-to-separation ratio, the parameter $\omega$ is found by equating the coefficient of the quadratic term with $-m\omega^2/4$~\cite{QuantGrav-Ankit2022}.
For more than one central interaction present simultaneously, the ``total'' $\omega$ is given by a Pythagoras-like sum for the individual interactions: $\omega^2 =  
\sum_i \omega_i^2$~\cite{QuantGrav-Ankit2022}.
Note that the covariance matrix is entirely determined by frequencies $\omega_0$ and $\omega$.
For the standard Newtonian gravity one finds $\omega_N^2 = 4Gm/L^3$, and for the MONDian gravity described by Eq.~\ref{eq:ForceIdenticalMasses} we obtain
\begin{equation}
\omega^2_M = \frac{8}{3}(\sqrt{2}-1) \frac{\sqrt{Gma_0}}{L^2}.
\end{equation}

In an actual experiment, the gravitational force is accompanied by non-gravitational interactions, which also admit an entangling effect. The most important of these is the Casimir attraction between the two masses. Following the algorithm described above, we find that contribution to entanglement is parameterized by~\cite{Casimir-Emig2007,PhysRev.73.360} (see Supplemental Material for complete derivation)
\begin{equation}
\omega_{C}^2 = \frac{2\hbar c R_0^6}{m\pi L^9} \sum_{n=0}^{9} C_n (n+7)(n+8) \frac{R_0^n}{L^n},
\end{equation}
where the constants $\qty{C_n}$ were determined in Ref.~\cite{Casimir-Emig2007} and $R_0$ stands for the radius of each metallic sphere.
Placing the two objects in close proximity increases gravitational coupling, but such a setup would also encounter an overwhelmingly large Casimir attraction, practically suppressing all gravitational effects.
On the other hand, increasing the separation to an extent that the Casimir forces are negligible also diminishes the gravitational coupling drastically, leading to a large amount of time required to accumulate a detectable entanglement.
We therefore propose a configuration where the two masses are separated by a moderate distance such that the force gradients of gravitational and Casimir interactions are comparable.

In Fig.~\ref{fig:Entanglement} we show the entanglement gain between two Platinum spheres\footnote{Although Osmium is the densest naturally occurring material, it is hard and brittle, making it difficult to shape as a sphere. Platinum is almost as dense and malleable, and it can be forged into a spherical shape of a predetermined size.}
of radius $R_0 = 10$ $\mu$m separated by $L = 10R_0$. 
The configuration admits an internal \emph{relative acceleration} of $\approx a_0/100$, readily in the regime of our interest and beyond the limits tested with torsion pendulums experiments~\cite{TestNewton-Little2014}. 
The entanglement accumulation due to the MOND, Newtonian gravity, and the Casimir interaction are independently characterized by 
$(\omega_M,\omega_N,\omega_C) = (3.06,1.55,3.53) \times 10^{-4}$.
The density of Platinum is $21.45$ g/cm$^3$, which implies the considered objects have a mass $m \approx 90$ ng.
We propose to prepare their initial (Gaussian) states by cooling the masses in harmonic traps of frequency $10$ MHz, which results in an initial ground state localization of  $\sqrt{\hbar/2m\omega_0} \approx 0.1$ fm, achievable with currently available optical tweezers.
The left panel of Fig.~\ref{fig:Entanglement} shows the ideal case of $\bar{n} = 0$, where the masses are cooled to the ground state. The middle panel shows a realistic scenario where the initial state is a mixed Gaussian with a no- zero phonon number $\bar n = 0.0015$. 
The covariance matrix for initially uncorrelated thermal states, characterized by the average phonon number $\bar{n}$, is given by $\qty(2\bar{n}+1)$ times the one obtained for the initial ground states.
The solid lines represent the entanglement generated by the simultaneous action of gravitational and Casimir forces, while dashed lines serve as a reference, illustrating the partial contributions of individual interactions.
The MONDian entanglement is much higher than what is accumulated by Newtonian gravity.
Note that the shaded region within the two curves corresponds to other possible approaches to modified gravity predicting a force gradient between Newtonian and MONDian.
Hence, the experimenter is not required to quantify entanglement, and a mere entanglement witnessing within the time window $110-125$ milliseconds will certify departures from canonical gravity. 
The right panel of Fig.~\ref{fig:Entanglement} shows this experimental time window for initial states with different phonon numbers.

\emph{\textbf{Discussion.}}
In order to experimentally estimate the amount of entanglement one needs to measure elements of the covariance matrix, which in principle requires finding position and momentum correlations~\cite{Aspelmeyer-CavOpt-Rev}. The question emerges as to whether a simpler method could exist to detect the stronger-than-Newtonian force, e.g., by measurements of position only. This is of course possible, but it turns out that the average displacements of the masses are about three orders of magnitude smaller than their standard deviations, at least for the parameters considered here.
While the alternative models give rise to larger than Newtonian variances, from a practical perspective one needs to ensure that the predicted increment is not mimicked by the noise in the system.
The added value of entanglement witnessing is its independence from noise characterization, whereas the added complexities are not high since the measurements of different mechanical quadratures can be accomplished, e.g., by choice of local phase in homodyne detection~\cite{Aspelmeyer-CavOpt-Rev}.

\emph{Environmental Decoherence:}
We now estimate the effects of decoherence. The interaction of the considered masses with ambient thermal photons and air molecules suppresses coherences over long distances. For the gravitational entanglement proposals, such decoherence mechanisms have been studied in Refs.~\cite{QuantGrav-Bose2017,QuantGrav-Vedral2017,QuantGrav-Tanjung2020,QuantGrav-Decoherence-Rijavec2021,QuantGrav-Animesh2021}. 
In the regime where the superposition is much smaller than the
wavelength of the scattering particles,
the coherence time for a quantum superposition across a distance $\Delta x$ is given by 
$\tau = \frac{1}{\Lambda (\Delta x^2)}$, 
where $\Lambda$ depends on the characteristics of environment. The interactions with ambient thermal photons is quantified by $\Lambda_{\text{ph}} \approx 10^{36} R_0^6 T_{\text{env}}^9$~\cite{book-Decoherence-Maximilian}.
In the discussed setup, we take the final width of the wave packet at time 125 ms (upper limit of time window at $\bar n = 0.0015$), i.e., $\Delta x \approx 0.76$ nm.
An ambient temperature of 2 mK (achievable with ${}^3$He/${}^4$He dilution refrigerator~\cite{Dilution-Referigerator}) implies that the coherence time is $\tau_\text{ph} \sim 10^{36}$ s. This is not surprising as the intensity of the thermal background peaks at $\approx 1.45$ m, which is nine orders of magnitude larger than the position spread.
Interactions with ambient air molecules is still an issue. 
At 2 mK ambient temperature the thermal de Broglie wavelength of air molecules is an order of magnitude larger than $\Delta x$, and hence the decoherence is quantified by~\cite{book-Decoherence-Maximilian}
\begin{equation}
\Lambda_\text{air} = \frac{8 n_\text{air}}{3\hbar^2}  \sqrt{2\pi m_\text{air}} (k_B T_\text{env})^{3/2}  R_0^2,
\end{equation}
where $m_\text{air} \approx 0.5 \times 10^{-25}$ kg is the average mass of an air molecule, and $n_\text{air}$ is the air density in the vacuum chamber.
In an ultra-high vacuum with a particle density $\approx 3 \times 10^{4}$ cm$^{-3}$~\cite{url-UHV}, we get a coherence time $\approx$ 0.94 s, well above the requirements of this proposal. Note the short experimental time requires a free fall inside a vacuum chamber of height $\lesssim 10$ cm, adding to the ease of creating and maintaining low pressures and temperatures. All this suggests a plausible, though demanding, terrestrial experiment. Accordingly, in the following section we further address the effects of Earth's proximity.

\emph{Tabletop Experiment:}
\begin{figure}[!b]
\centering
\includegraphics[width=\linewidth]{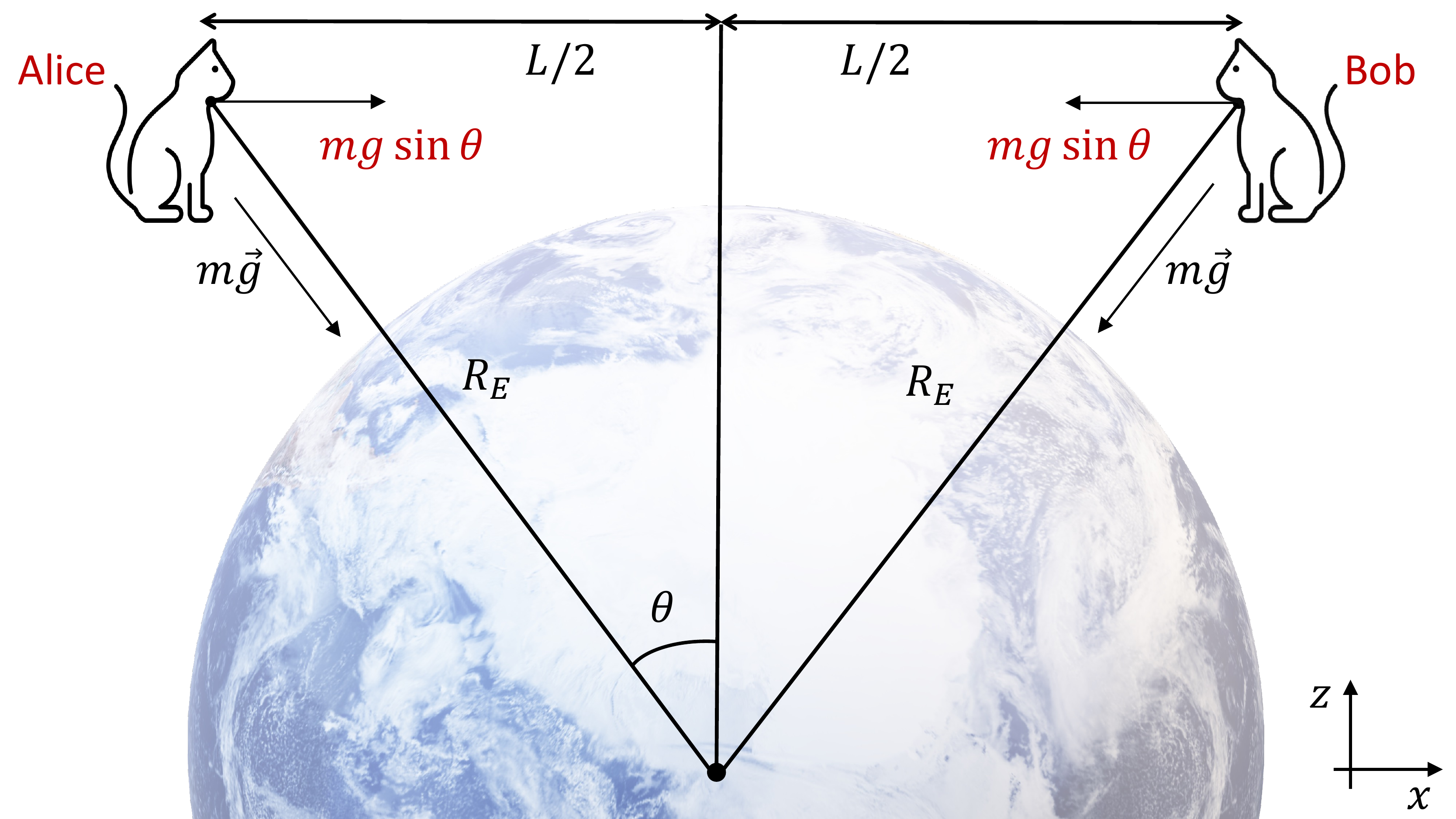}
\caption{Origin of tidal forces in a bipartite system of two identical masses $m$ separated by a distance $L$, next to the surface of the earth (extremely not to scale). Here $g = |\vec{g}| = 9.81$ m/s$^2$ is the terrestrial acceleration, and $R_E = 6371$ km denotes the radius of the earth. The distances are over-exaggerated as in fact $L \sim  10^{-11}R_E$. Yet, the tidal components along the line joining the masses are two orders of magnitude larger than mutual gravity.}
\label{fig:TidalDepiction}
\end{figure}
Next to the surface of the Earth each particle experiences a strong pull towards the center of the planet, which contributes non-zero components along the line connecting the centers of the masses. These tidal forces arising due to the non-uniform gravity of the earth are illustrated in Fig.~\ref{fig:TidalDepiction}. 
Since that the acceleration due to gravity is $\sim 10^{11}a_0$, we treat the tidal forces 
as
horizontal components of the Newtonian force $mg$, which add linearly to the mutual gravity.
As seen in the figure, the magnitude of the tidal forces is $mg \sin\theta$, and they drift the masses towards each other. Replacing $\sin\theta$ using geometry in Fig.~\ref{fig:TidalDepiction} we find:
\begin{eqnarray}
\vec{F}^\text{(Tidal)}_{A} &=&  - m\frac{g}{R_E} \qty( - \frac{L}{2} + x_A )  \ \hat e_x,
\nonumber    \\
\vec{F}^\text{(Tidal)}_{B} &=& - m\frac{g}{R_E} \qty( \frac{L}{2} + x_B ) \ \hat e_x,
\end{eqnarray}
where $\hat e_x$ is the unit vector along $x$ axis (origin at the center between the masses), and $x_A(x_B)$ is the displacement of mass $A(B)$ from its initial position.
The tidal potential is therefore
\begin{eqnarray}
V^\text{(Tidal)} &=&  \frac{1}{2} m  \frac{g}{R_E} \left[ \qty(-\frac{L}{2}+ x_A)^2
+ \qty(\frac{L}{2}+ x_B)^2  \right],
\end{eqnarray}
which appears to confine each mass in a harmonic trap of frequency $\omega_E = \sqrt{g/R_E} \approx 0.00124$.
Even though $\omega_E$ is an order of magnitude larger than the characteristic frequencies of mutual gravity (and the tidal acceleration about hundred times that of mutual attraction), it is a \emph{local} interaction and as such it cannot generate any entanglement.
Yet, it does influences the wave function locally, which in turn can speed up / slow down the entanglement accumulation due to mutual interactions. 
To check if this effect is relevant we solve the time evolution of covariance matrix with mutual interactions truncated at the quadratic term and with the tidal potential.
The solution reads:
\begin{widetext}
\begin{eqnarray}
\bm{\sigma}_{00}(\bm{\sigma}_{02})
&=&	\frac{\hbar}{4m\omega_{0}} \qty{
\qty( \cos^{2}(\omega_{E}t)+\frac{\omega_{0}^{2}}{\omega_{E}^{2}}\sin^{2}(\omega_{E}t) )
+\!(\!-\!) \
 \qty[ \cos^2\qty(\sqrt{\omega_E^2-\omega^2} \, t) + \frac{\omega_0^2}{\omega_E^2-\omega^2} \sin^2\qty(\sqrt{\omega_E^2-\omega^2} \, t) ]
}, 
\nonumber \\
\bm{\sigma}_{11}(\bm{\sigma}_{13})
&=& \frac{1}{4}m\hbar\omega_{0} 
\qty{
\qty( \cos^{2}(\omega_{E}t)+\frac{\omega_{E}^{2}}{\omega_{0}^{2}}\sin^{2}(\omega_{E}t) )
+\!(\!-\!) \
\qty[ \cos^2\qty(\sqrt{\omega_E^2-\omega^2} \, t) + \frac{\omega_E^2-\omega^2}{\omega_0^2} \sin^2\qty(\sqrt{\omega_E^2-\omega^2} \, t) ]
},
\nonumber \\
\bm{\sigma}_{01}(\bm{\sigma}_{03})
&=&	\frac{\hbar}{8}  
\Bigg[
\Big(\frac{\omega_{0}}{\omega_{E}}-\frac{\omega_{E}}{\omega_{0}}\Big)\sin(2\omega_{E}t)
+\!(\!-\!) \
\qty( \frac{\omega_0}{\sqrt{\omega_E^2-\omega^2}}  - \frac{\sqrt{\omega_E^2-\omega^2}}{\omega_0} ) \sin\qty(2\sqrt{\omega_E^2-\omega^2} \, t)
\Bigg],
\label{eq:CovMat_FreeFall_Tidal}
\end{eqnarray}
\end{widetext}
where again $\omega_0$ is the frequency of harmonic trap used for preparing initial Gaussian states, and $\omega$ encodes all mutual interactions between the two masses. While this solution looks distinct from that in Eq.~\eqref{eq:CovMat_FreeFall}, the entanglement negativity in Fig.~\ref{fig:Entanglement} decreases relatively by a negligible amount $\sim 10^{-9}$, many orders of magnitude smaller than the achievable laboratory precision. In short, for the parameters considered in this work, tidal forces are insignificant in entanglement dynamics.

\emph{\textbf{Summary.}}
We proposed an entanglement experiment 
with the potential outcome revealing non-Newtonian gravity at small accelerations. 
It was motivated by modified gravity models, and the concrete calculations were performed using MOND parameters. 
Quantum particles naturally source weak gravity and small accelerations and provide an alternative route for testing modified gravity. The entanglement measurement brings independence from detailed noise characterization while adding only moderate complexity. Calculations including Casimir forces, tidal effects, and environmental decoherence reveal the possibility of a tabletop experiment.

\emph{\textbf{Acknowledgments.}}
This work is jointly supported by 
(i) Xiamen University Malaysia,
via projects \sloppy{XMUMRF/2022-C10/IPHY/0002} and \sloppy{XMUMRF/2021-C8/IPHY/0001},
(ii) SERB-DST, Govt.~of India, via project no. \sloppy{CRG/2022/009359}, and
(iii) project 2021/43/P/ST2/02141 co-funded by the NCN and the EU's Horizon 2020 under the Marie Sk\l{}odowska-Curie grant agreement 945339. 
A.K. is supported by the (i) Foundation for Polish Science through the IRAP project ``International Centre for Theory of Quantum Technologies'' (contract no. 2018/MAB/5) co-financed by EU's Smart Growth Operational Program, and 
(ii) QuantERA II project ``Experiment and Theory of Resources in Quantum Technologies'' (contract no. 2021/03/Y/ST2/00178) that has received funding from EU’s Horizon 2020. We gratefully thank the anonymous reviewer for their thorough reading and attention to to detail; their insightful comments resulted in a significant improvement in the clarity, correctness, and completeness of this work. 
A.K. thanks Rishabh Duhan (TIFR, India) for explaining the current technological limits for creating and maintaining low pressures and temperatures.
We acknowledge National Supercomputing Mission (NSM) for providing computing resources of `PARAM Ganga' at Indian Institute of Technology Roorkee, which is implemented by C-DAC and supported by the Ministry of Electronics and Information Technology (MeitY) and Department of Science and Technology (DST), Government of India.


%

\cleardoublepage
\begin{center}
\LARGE
\underline{SUPPLENMENTAL MATERIAL}
\end{center}

\section*{Different MOND formulations}

The generalizations of Newton's second law of motion and Poisson's equation describe a modification to the acceleration $\vec{a}$ of an object of mass $m$ under the influence of a non-gravitational force $\vec{F}$ and gravitational force $-m{\nabla}\Phi$, where $\Phi$ is determined via the solution of a set of potentially modified Poisson equations for the set of potentials $\qty{\phi_{i}}$ [see main text for details]. 
Examples of several specific proposals are given in Table~\ref{tab:ManyMOND}, where non-zero $\mu_{ij}$ and $G_{i}$ are detailed.
We note that there can additionally exist formulations of MOND where the quantity $a_{0}$ itself may depend on $\Phi$ or $|\nabla \Phi|$~\cite{MOND2Body-Zhao2010}, where $\Phi$ appears as a `mass term' in the field equations~\cite{RMOND-Zlosnik2021}, or where terms involving higher derivatives of potentials appear in the field equations~\cite{MOND-General-Milgrom2023}.

\begin{table}[H]
\centering
\caption{Different formulations of the MOND theory.}
\begin{tabular}{ l|c|c }
\toprule
Model & Potentials & Properties \\
\midrule\midrule
Modified Inertia~\cite{MOND-Milgrom1992}	& $\phi_1 = \Phi$ & 
$\nu=\nu(|\vec{a}|/a_{0})$ 	\\
&	&	$\mu_{11}=1$
\\	\midrule
AQUAL~\cite{Lag4MOND-Bekenstein1984} & $\phi_1 = \Phi$	& 
$\nu=1$	\\
&	&	$\mu_{11}=\mu_{11}(|\nabla\Phi|/a_{0})$ \\
& & $G_{1}=G$
\\	\midrule
TeVeS~\cite{TeVeS-Bekenstein2004} & $(\phi_1, \phi_2) = $ &	
$\nu=1$				\\
& $(\Phi - \varphi,\varphi)$	&	$\mu_{11} = 1$	\\
&	&	$\mu_{22} =\mu_{22}(|\nabla\varphi|/a_{0})$
\\	
& & $G_{1}=G_{2}=G$\\
\midrule
QUMOND~\cite{MOND-QLinear-Milgrom2009} & $(\phi_1,\phi_2) = $ &
$\nu=1$				\\
& $(\Phi,\varphi)$	&	$\mu_{11} = 1$	\\
&	&	$\mu_{12} =\mu_{12}(|\nabla\varphi|/a_{0})$	\\
&	&	$\mu_{22} = 1$	\\
&	&	$G_{2}=G$
\\	\midrule
TRIMOND~\cite{Milgrom:2023idw} & $(\phi_1,\phi_2,\phi_3) =$ &
$\nu=1$				\\
& $(\Phi,\varphi,\psi)$	&	$\mu_{13} = 1$	\\
&	&	$\mu_{22} =\mu_{22}(\nabla\varphi,\nabla\psi)$	\\
&	&	$\mu_{23} =\mu_{23}(\nabla\varphi,\nabla\psi)$	\\
&	&	$\mu_{31} = 1$	\\
&	&	$\mu_{32} =\mu_{32}(\nabla\varphi,\nabla\psi)$	\\
&	&	$\mu_{33} =\mu_{33}(\nabla\varphi,\nabla\psi)$	\\
&	&	$G_{1}=G$ 		
\\  \bottomrule
\end{tabular}
\label{tab:ManyMOND}
\end{table}

\section*{Bipartite covariance matrix}
\label{sec:CovMat}

The covariance matrix formalism is based on the first two statistical moments of a quantum state. Given a bipartite system $AB$ with the displacement and momentum quadratures $ \hat u = ( \hat x_A, \hat p_A, \hat x_B, \hat p_B)^T$, the (symmetric) covariance matrix is defined by~\cite{Negativity-Vidal2002,Negativity-CV-Serafini2004,Negativity-Gaussian-Adesso2005}:
\begin{equation}
\bm{\sigma}_{jk} = \bm{\sigma}_{kj}  = \frac{1}{2} \ev{ \acomm{\hat u_j}{\hat u_k} }  -  \ev{ \hat u_j }\ev{ \hat u_k } .
\end{equation}
In the block form it reads
\begin{equation}
    \bm{\sigma} \equiv \mqty(
\bm{\alpha} & \bm{\gamma} \\
\bm{\gamma}^T & \bm{\beta}
),
\end{equation}
where $\bm{\alpha}(\bm{\beta})$ contains the local mode correlation for $A(B)$, and $\bm{\gamma}$ describes the intermodal correlation. 
In this work we deal with two identical masses prepared in identical initial states, which implies that the local modes are same at all times: $\bm{\alpha} = \bm{\beta}$.

\section*{The Casimir interaction}

The Casimir interaction is central, and in the relevant limit of perfect metals the energy of attraction between two identical spheres is given by~\cite{Casimir-Emig2007,PhysRev.73.360}:
\begin{equation}
V_C(r) = - \frac{\hbar c R_0^6}{\pi(L+r)^7} \sum_{n=0}^{\infty} C_n \qty(\frac{R_0}{L+r})^n ,
\end{equation}
where $R_0$ is the radius of the sphere and the constants for the first ten terms were determined in Ref.~\cite{Casimir-Emig2007}:
\begin{eqnarray}
C_0 &=& + 143/16,
\nonumber	\\
C_1 &=& 0,
\nonumber	\\
C_2 &=&  +  7947/160,
\nonumber	\\
C_3 &=&  + 2065/32,
\nonumber	\\
C_4 &=&  + 27 \ 705 \  347 / 100  \  800,
\nonumber	\\
C_5 &=& - 55  \  251 / 64,
\nonumber	\\
C_6 &=&  + 1 373  \ 212  \ 550 \  401 / 144 \  506 880,
\nonumber	\\
C_7 &=& - 7  \ 583  \ 389 / 320,
\nonumber	\\
C_8 &=& - 2  \ 516 \  749  \ 144  \ 274  \ 023 / 44 \  508 \  119 \  040,
\nonumber	\\
C_9 &=&  + 274  \ 953  \ 589  \ 659 \  739 / 275  \ 251  \ 200.
\end{eqnarray}
While the series, in principle, does not converge for any value of $R_0/L$, the results presented in Fig.~1 of Ref.~\cite{Casimir-Emig2007} show that the leading term alone is sufficient to describe the configuration considered in this work. Nevertheless, for possible implementation in other configurations, we expand all available terms in binomial series of $r/L$:
\begin{eqnarray}
V_C(r) &\approx& - \frac{\hbar c R_0^6}{\pi} \sum_{n=0}^{9} C_n \frac{R_0^n}{L^{n+7}} \qty( 1 + \frac{r}{L} )^{-(n+7)}    \nonumber   \\   &=& - \frac{\hbar c R_0^6}{\pi} \sum_{n=0}^{9} C_n \frac{R_0^n}{L^{n+7}} \sum_{k=0}^{\infty} \binom{-(n+7)}{k} \frac{r^k}{L^k}   \nonumber   \\ 
&=& - \frac{\hbar c R_0^6}{\pi L^7} \sum_{n=0}^{9} C_n \frac{R_0^n}{L^n} \sum_{k=0}^{\infty} (-1)^k \binom{n+k+6}{k} \frac{r^k}{L^k}. \hspace*{1cm}
\end{eqnarray}
Using the algorithm developed in Ref.~\cite{QuantGrav-Ankit2022} and discussed in the main text, we find that the Casimir energy truncated at quadratic term ($k=2$) gives the contribution to entanglement parameterized by
\begin{equation}
\omega_{C}^2 = \frac{2\hbar c R_0^6}{m\pi L^9} \sum_{n=0}^{9} C_n (n+7)(n+8) \frac{R_0^n}{L^n}.
\end{equation}

\end{document}